# A Framework for Data-Based Turbulent Combustion Closure: *A Priori* Validation

Rishikesh Ranade, Tarek Echekki[1]

*Department of Mechanical and Aerospace Engineering, North Carolina State University, Raleigh, NC 27695-7910, USA*

**Abstract**

Experimental multi-scalar measurements in laboratory flames have provided important databases for the validation of turbulent combustion closure models. In this work, we present a framework for data-based closure in turbulent combustion and establish an *a priori* validation of this framework. The approach is based on the construction of joint probability density functions (PDFs) and conditional statistics using experimental data based on the parameterization of the composition space using principal component analysis (PCA). The PCA on the data identifies key parameters, principal components (PCs), on which joint scalar PDFs and conditional scalar means can be constructed. To enable a generic implementation for the construction of joint scalar PDFs, we use the multi-dimensional kernel density estimation (KDE) approach. An *a priori* validation of the PCA-KDE approach is carried out using the Sandia piloted jet turbulent flames D, E and F. The analysis of the results suggests that a few PCs are adequate to reproduce the statistics, resulting in a low-dimensional representation of the joint scalars PDFs and the scalars' conditional means. A reconstruction of the scalars' means and RMS statistics are in agreement of the corresponding statistics extracted directly from the experimental data. We also propose one strategy to recover missing species and construct conditional means for the reaction rates based on a variation of the pairwise-mixing stirred reactor (PMSR) model. The model is demonstrated using numerical simulations based on the one-dimensional turbulence (ODT) model for the same flames.

***Keywords*:** data-based modeling; joint probability density function; principal component analysis; kernel density estimation.

---

[1] Corresponding Author. Address: Department of Mechanical and Aerospace Engineering, North Carolina State University, 911 Oval Drive, Campus Box 7910, Engineering Building III, Room 3252, Raleigh, NC 27695-7910, USA. Fax: +1 919 515 7968, E-mail address: techekk@ncsu.edu (T. Echekki).

# 1. Introduction

In turbulent combustion modeling, the complex interaction between chemistry and turbulence results in unclosed chemical source terms. Several methodologies (e.g. the flamelet model [1], conditional moment closure model [2]) have been used to build models that close these highly non-linear source terms. The bulk of these methodologies are designed to capture 2 principal ingredients: 1) an adequate description of the composition space, which seek to relate the different variables in the thermo-chemical scalars' vector, and 2) an adequate description of the statistical distribution of these variables in the form of a joint probability density functions (PDF). A principal challenge in capturing the composition space is the ability to reduce the description of the variables of the thermo-chemical scalars' vector in terms of a limited set of parameters (e.g. mixture fraction, progress variable) that can adequately describe this space. These parameters equally are useful in determining a simplified description of the joint PDF of the thermo-chemical scalars. For example, under the steady and adiabatic flamelet assumption, the description the composition space may be parameterized in terms of the mixture fraction and the scalar dissipation rate. Presumed shape PDFs in terms of these parameters or related parameters may be adequate to address relatively complex flows where the flamelet assumption applies.

However, there are many instances where the pre-selection of parameters for the composition space is not trivial. These include the presence of non-equilibrium effects, such as extinction and reignition, the presence of multiple streams for mixing or multiple modes of combustion (e.g. premixed and non-premixed) or the presence of heat losses near boundaries. Under these instances, models for the high-dimensional joint PDFs of the thermo-chemical scalars become invariably complex. Such models include the solution for the joint PDF transport equations [3]. A common strategy with these models is the construction of an equivalent system of interacting notional particles that develop with their own chemistry, which also track the evolution of the reacting composition space. This method has been successfully demonstrated in the past for a wide range of problems considering finite rate chemistry effects (see for example, Refs. [4-9]). However, the transport of a joint PDF comes at a potentially high computational cost for complex chemical systems and large computational problems where a large number of notional particles must be retained for an accurate evaluation of the PDFs. Alternative strategies to alleviate the potential cost of a joint PDF transport equation have been proposed, including strategies to parameterize the composition space (e.g. by adopting a flamelet assumption) or via the multi-environment PDF method [10].

The two limits of a presumed PDF shape with a limited set of parameters and the solution for a PDF transport equation leave ample room in between for intermediate approaches, which may be based on the construction of PDFs based on simulation [11-14] or experimental data. For example, Goldin and Menon [11, 12], Sankaran et al [13] and Calhoon et al [14] built joint scalar PDFs, which are parameterized by an appropriate set of lower moments using simulations based on the linear-eddy model (LEM) [15]. The PDFs are generally trained on simpler canonical flows, such as scalar decay in homogeneous turbulence or co-flowing jet configurations. The LEM is a 1-D model that can accommodate any degree of chemical complexity based on the coupling of reaction and diffusion processes in a deterministic fashion and a stochastic implementation for turbulent stirring. As the model is implemented in physical space, it is capable of generating statistics from which a description of the composition space and its parameterization and the joint thermo-scalars PDFs can be constructed. Extensions of this modeling approach are the one-dimensional turbulence (ODT) model [16] and the more recent hierarchical parcel-swapping (HiPS) model [17], both developed by Kerstein.

The idea of constructing statistics from low-dimensional stochastic simulations also can be extended to single-point (or line) multi-scalar experimental measurements. In these measurements, major species and temperature are measured in multiple shots at different points of the flow. An example of an excellent resource for such measurements are the ones based on the Turbulent Nonpremixed Flames (TNF) Workshop (https://www.sandia.gov/TNF/abstract.html). While the flames measured for the TNF Workshop have been primarily the testing beds for turbulent combustion closure models, it is easy to perceive the potential of such measurements in practical flows where a simple description of the statistics can no longer be represented in terms of the usual set of prescribed parameters.

If we are to rely on multi-scalar experimental data to construct closure models (i.e. an adequate description of the composition space and the statistical distributions), what strategies should be adopted? First, we would like to be able to, again, construct a minimum set of parameters that can be used to reconstruct the composition space; conditional means of the thermo-chemical scalars and their chemical source terms can be determined in terms of these parameters. Second, the joint PDF will be constructed based on these parameters with the assumption that a lower dimensional joint PDF can be constructed.

In this work, we propose a novel framework to construct closure models for turbulent combustion based single-point multi-scalar laser-based measurements in different regions of the reacting flow. Then, we attempt to demonstrate a key component of this framework by efficiently constructing joint PDFs and conditional statistics based on the experimental data and a dimensionality reduction of the thermo-chemical space. This dimensionality reduction is carried out using principal component analysis (PCA) [18]. PCA generates a map from the thermo-chemical scalars' vector to a smaller vector of principal component (PCs). The mapping can be reversed to determine temperature and species composition based on the PCs. The PCs, then, serve as generic substitutes for the common conditioning parameters in turbulent combustion (e.g. mixture fraction, progress variable, scalar dissipation rates). Invariably, there are alternative reduction methods to PCA, including non-linear forms of PCA. However, we have found that PCA to be an effective and optimal tool to reduce complex composition space.

A second demonstration is related to the construction of a joint PDF conditioned on a small and finite number of PCs. The construction is based on the kernel density estimation (KDE) [19] method. KDE is a data-based PDF construction method that can accommodate a hierarchy of PDF shapes and dimensions. In a recent study, Miles and Echekki [20] demonstrated the use of KDE for the construction of the marginal uni-variate PDFs of the mixture fraction and temperature in LES using data from ODT simulations of Sandia jet diffusion flames D, E and F [21]. The resulting PDFs were able to exhibit bi-modal shapes during extinction and reignition events in the Sandia flames.

By integrating the joint PDFs with the conditional statistics of the thermo-chemical variables, mean and RMS statistics of these variables can be derived. We propose the PCA-KDE approach to construct thermo-chemical scalars' statistics as a first step towards the development of closure models for different combustion problems based on experimental data that is specific (or at least closely related) to those problems.

The PCA-KDE approach is demonstrated using experimental data for the well-characterized piloted Sandia flames D, E and F [21]. These data sets contain instantaneous point measurements of thermo-chemical scalars including temperature and $H_2$, $O_2$, OH, $H_2O$, $CH_4$, CO and $CO_2$ at different axial and radial locations. PDFs conditioned on a finite number of PCs are constructed at each location from these instantaneous measurements. Moreover, conditional statistics based on PCs are constructed for thermo-chemical scalars from an ensemble of data available from the three flames. The PDFs computed at each location for each flame is integrated with the conditional statistics to determine the Favre and conditional averages and RMS statistics of thermo-chemical scalars. In Sec. 2, we present the proposed framework and underlying theory. Then, the results of the *a priori* study of the framework are presented and discussed in Sec. 3 based on the Sandia flames D, E and F results. Finally, we summarize our observations and propose strategies to complete the proposed framework in Sec. 4.

## 2. Proposed Framework

### 2.1. Model Formulation

The proposed framework consists of two primary steps: 1) a pre-processing step in which PCs are identified based on data and the joint PDFs and conditional means, which are parameterized based on the PCs, are constructed, and 2) the solution of the averaged or filtered PCs within the context of Reynolds-averaged Navier-Stokes (RANS) or large-eddy simulation (LES) frameworks. Therefore, either RANS or LES will involve the solution for the transport equations of PCs instead of the usual parameters adopted for different combustion problems (e.g. mixture fraction mean and variance, progress variable). In this study, we address the first step to establish whether PCs can be constructed to develop conditional means for the thermo-chemical scalars and joint PDFs based on the PCs and then reconstruct the means and RMS profiles of the thermo-chemical scalars. A follow-up study will address in detail the second step, which involves the development of closure terms for the PCs equations and the combustion problem with transport of the PCs. For the purpose of illustrating the model, we use a standard RANS model for the solution of the mixture density, the linear momentum and the PCs:

- **Continuity:**
$$\frac{\partial \bar{\rho}}{\partial t} + \frac{\partial \bar{\rho}\tilde{u}_j}{\partial x_j} = 0 \tag{1}$$

- **Momentum:**
$$\frac{\partial \bar{\rho}\tilde{u}_i}{\partial t} + \frac{\partial \bar{\rho}\tilde{u}_i\tilde{u}_j}{\partial x_j} = -\frac{\partial \bar{p}}{\partial x_i} + \frac{\partial}{\partial x_j}\left[2\bar{\rho}(\nu_T + \nu)\tilde{S}_{ij}\right], \quad i = 1,2,3. \tag{2}$$

- **PCs:**
$$\frac{\partial \bar{\rho}\tilde{\phi}_k}{\partial t} + \frac{\partial \bar{\rho}\tilde{u}_j\tilde{\phi}_k}{\partial x_j} = \frac{\partial}{\partial x_j}\left[\bar{\rho}\left(\frac{\nu_T}{\mathrm{Sc}_T} + \nu_k\right)\frac{\partial \tilde{\phi}_k}{\partial x_j}\right] + \bar{s}_k, \quad k = 1,\cdots,N_{\mathrm{PC}} \tag{3}$$

In the above expressions, the symbols "¯" and "~" correspond to Reynolds averaging and density weighted averaging, respectively. $\bar{\rho}$ is the mean density, $\tilde{u}_i$ and $\tilde{\phi}_k$ are the density-weighted velocity component in the $i$th direction and the $k$th PC. There are $N_{\mathrm{PC}}$ PCs that are retained to transport the solution within RANS. From this solution, averaged thermo-chemical scalars can be reconstructed based on the procedure outlined below. According to the governing equations, the PCs are not passive scalars; they do carry chemical source terms, $\bar{s}_k$. The modeling of these source terms is one of the principal closure requirements for the implementation of the model. This closure is only addressed briefly here. While the scope of the present study is to demonstrate that a meaningful definition for the PCs can be extracted from experimental data and that thermo-chemical scalars can be constructed based on knowledge of the PCs. In the above equations, $v_T$ is a turbulent kinematic viscosity, $\text{Sc}_T$ is a turbulent Schmidt number and $S_{ij}$ is the *ij* component of the rate-of-strain tensor. Additional closure to model the turbulent viscosity also is needed, such as the use of a k-ε model or a Reynolds stress model. One may argue that the transport of the PCs variances in an analogy with typical moment-based methods may be needed as well. Below and based on the *a priori* validation, we will argue that the transport of such variances may not be needed to advance the solution of the PCs.

The set of PCs selected is not unique. They must be derived from some knowledge of the accessed composition space of interest by relying on resolved multi-scalar point measurements. In the present study, we will derive PCs and propose strategies for recovering statistics for temperature and the mixture composition based on the PCs transport equations. The PCs are constructed on instantaneous single-shot multi-scalar measurements to determine 1) the required number of PCs that adequately represents the composition space and 2) the necessary mapping relations from thermo-chemical scalars to PCs and *vice versa*. From this information, two sets of statistics can be constructed:

- **Conditional means of thermo-chemical scalars:** These means for temperature and measured species mass fractions, $\langle \boldsymbol{\theta}|\boldsymbol{\phi}\rangle = (\langle T|\boldsymbol{\phi}\rangle, \langle Y_1|\boldsymbol{\phi}\rangle, \langle Y_2|\boldsymbol{\phi}\rangle, \ldots, \langle Y_{N-1}|\boldsymbol{\phi}\rangle)$ are averaged conditioned on the values of the retained PCs, $\boldsymbol{\phi}$. These means constitute what is equivalent to generalized flame libraries in combustion. The conditional means are obtained from the single-shot multi-scalar measurements.

- **Joint scalar PDFs:** Scalar PDFs, $P(\boldsymbol{\phi}; \tilde{\boldsymbol{\phi}})$, are constructed on each position of the point measurements and are conditioned upon the Favre-means of the retained PCs as outlined below. Strategies to pool data from different points in the measurement field to construct a multi-dimensional PDF also are implemented as discussed below.

The Favre means of a given *k*th thermo-chemical scalars, $\tilde{\theta}_k$, can be evaluated as follows:

$$\tilde{\theta}_k(\tilde{\boldsymbol{\phi}}) = \frac{\int \langle \rho | \boldsymbol{\phi} \rangle \langle \theta_k | \boldsymbol{\phi} \rangle P(\boldsymbol{\phi}; \tilde{\boldsymbol{\phi}}) d\boldsymbol{\phi}}{\bar{\rho}}, \text{ where } \bar{\rho} = \int \langle \rho | \boldsymbol{\phi} \rangle P(\boldsymbol{\phi}; \tilde{\boldsymbol{\phi}}) d\boldsymbol{\phi} \qquad (4)$$

This relation recovers the Favre-averaged thermo-chemical scalars given a solution for the PCs vector $\tilde{\boldsymbol{\phi}}$. A similar relation can be used to model the PCs source term in Eq. (3):

$$\bar{s}_k(\tilde{\boldsymbol{\phi}}) = \int \langle s_k | \boldsymbol{\phi} \rangle P(\boldsymbol{\phi}; \tilde{\boldsymbol{\phi}}) d\boldsymbol{\phi} \qquad (5)$$

where $\langle s_k | \boldsymbol{\phi} \rangle$ is the conditional mean of the PCs source terms. As outlined below, the PCs instantaneous source terms can be derived starting with the thermo-chemical scalars' source terms and conditional means for these source terms can be determined the same way the conditional means of the thermo-chemical scalars. The closure for the PCs source terms is an integral component of the proposed framework and a necessary step to be able advance the solution for the system of equations (1) to (3). Experimental data, clearly, does not provide direct information about the chemical source terms of the measured species or temperature and hence they need to be reconstructed from the available information.

The reconstruction of chemical source terms must address several challenges. First, a description of the chemistry (detailed or reduced) must be available to evaluate the thermo-chemical scalars source terms (from which the PCs source terms can be determined as discussed below). Second, the measurements include only a subset of the thermo-chemical scalars' vector that is described by the available chemical mechanism. Therefore, either the missing species must be recovered, or a mechanism must be constructed by including only the measured species or a subset of theses measured species. Finally, noise from instantaneous experimental measurements, no matter how small it is relative to the actual signal, can be amplified through the non-linear dependence of the reaction rate on temperature and species concentrations.

To address these challenges, we will implement a combination of strategies that address potentially noisy (with experimental error) and partial (with missing species) data. We will illustrate one strategy here based on a minor variation of the pairwise-mixing stirred reactor (PMSR) model [22]. The approach is based on the following hypothesis: by locally (in composition space) mixing states based initially on single-short multi-scalar experimental measurements, we can recover missing species and establish smoother profiles for the measured and unmeasured species in composition space. The full thermo-chemical state available at the end of this iteration is used to construct instantaneous source terms and subsequently the conditional means of PC source terms. Although we are presenting only one strategy to reconstruct source terms for species and the PCs, combining this strategy with others that work to denoise the data can potentially improve the reconstruction of the reaction rates. We are currently investigating these strategies, including the use of PCA as well as deep machine learning techniques that have been demonstrated for a broad range of applications.

## *2.2. Chemical source term reconstruction using PMSR*

PMSR is a zero-dimensional stochastic reactor model in which a discrete set of particles, each with their own thermo-chemical composition, $\phi_i$, are integrated for a finite residence time, $\tau_r$ [22]. Within the context of our approach, these particles are assigned initially the thermo-chemical state of individual measurements. While, there are potentially alternative strategies to initialize the missing species, we chose to set these species concentrations to zero initially. During each time-step, these particles undergo two parallel events: (i) a deterministic reaction step based on a chemical mechanism and (ii) a stochastic mixing step based on interaction with other particles. The evolution of particle composition through reaction is described as follows:

$$\frac{d\theta_i}{dt} = S(\theta_i) \qquad (6)$$

where $S$ is the reaction rate of composition determined from a chemical mechanism. The evolution from mixing for a pair of particles (say, *a* and *b*) randomly chosen from the existing set is implemented as follows:

$$\frac{d\theta_i^a}{dt} = \frac{\theta_i^b - \theta_i^a}{\tau_m}, \qquad (7)$$

$$\frac{d\theta_i^b}{dt} = \frac{\theta_i^a - \theta_i^b}{\tau_m}. \qquad (8)$$

where $\tau_m$ is a mixing time, which is a prescribed fraction of the total residence time, $\tau_r$.

In our work, the modified PMSR model to reconstruct the chemical source terms of the experimentally measured temperature and species is implemented as follows. The instantaneous measurements at different radial and axial locations in the flames D, E and F are considered as the different particles in the PMSR model. A different PMSR is carried out at each measurement position to ensure local mixing, as proposed earlier. For the present validation studies based on methane-air mixtures, the 12-step augmented reduced mechanism (ARM) [23] is used to compute the reaction rates in Eq. (6). In the PMSR run, the particle composition state is initialized with the measured species and temperature while the rest of the species that are part of the chemical mechanism are set to zero. The PMSR is integrated up to a specified residence time. The particles are randomly paired with other particles in the reactor and the reshuffling of pairs is carried out at the end of each chemical timestep. In this work, we have considered several strategies for the determination of the residence time. One obvious strategy is to consider the measured OH, as a missing species and stopping the residence time when the OH concentration matches the target value. Alternatively, we carry out the PMSR simulation until a prescribed residence time is reached or if the measured species concentrations exceed the error bounds of the measurements. Based on comparisons of the different strategies, in the present study, we find that prescribing a residence time of $10^{-2}$ sec yields almost identical results to the setting a threshold for a target value for OH. Moreover, similar results are obtained for a range of residence times from $10^{-3}$ to $10^{-1}$ sec. Moreover, the temperature and measured species are curbed by switching off their mixing and reaction source terms if they evolve beyond the measured scalars' experimental uncertainty ranges. The recovery of the missing species of the chemical mechanism allows for the estimation of source terms of temperature and the measured species.

*2.3. Parameterization of the composition space with PCA*

As indicated earlier, we choose to parameterize the composition space using PCA. PCA is a statistical procedure that uses an orthogonal transformation to convert a set of observations of possible correlated variables into a set of values of linearly uncorrelated variables, the PCs. This method has been used previously to reduce the cost of turbulent combustion simulations by representing the large thermo-chemical space with a smaller set of PCs [18, 24-29]. PCA relates a vector $\boldsymbol{\theta} = (T, Y_1, Y_2, \ldots, Y_{N-1})$ of $N$ thermo-chemical scalars (here temperature and $N$-1 species mass fractions), which represent the composition space, and a vector $\boldsymbol{\phi} = (\phi_1, \phi_2, \ldots, \phi_N)$ of $N$ corresponding PCs using a linear relationship:

$$\boldsymbol{\phi} = \mathbf{Q}^\mathbf{T}\boldsymbol{\theta}, \qquad (9)$$

where $\mathbf{Q}$ is the $N \times N$ matrix of orthonormal eigenvectors of the correlation matrix of thermo-chemical scalars and the superscript "**T**" refers to the transpose of $\mathbf{Q}$. The transformation is defined in such a way that the PCs are ordered from highest to the lowest variance contributions to the data based on the magnitude of their eigenvalues. Based on our prior studies in turbulent combustion problems [24, 26], the first few PCs account for 95% and higher of the variability in the data and may be enough to represent to entire set thermo-chemical variables. Therefore, our modeling approach relies on the retention of a smaller number of PCs:

$$\boldsymbol{\phi}^{\mathrm{red}} = \mathbf{A}^\mathbf{T}\boldsymbol{\theta}. \qquad (10)$$

Here, the superscript "red" refers to the reduced set of $N_{\mathrm{PC}}$ PCs, which are retained to reconstruct the thermo-chemical scalars; while $\mathbf{A}^\mathbf{T}$ corresponds to the first set of columns of the matrix $\mathbf{Q}^\mathbf{T}$, which correspond to the $N_{\mathrm{PC}}$ retained PCs. The PCs serve a dual role in the proposed approach. First, they are used to parameterize the composition space by enabling a mapping from the retained PCs to the thermo-chemical scalars. This mapping represents an inversion of the relation in Eq. (2). It is implemented using a non-linear procedure based on artificial neural networks (ANN) regression [24-29]. We have found that the PCA-ANN mapping yields the same accuracy as mappings based on non-linear PCA methods. In the discussion below, and as implemented in Sec. 2.1, we will drop the superscript "red" and denote the vector of retained PCs as $\boldsymbol{\phi}$.

As derived by Sutherland and Parente [18], the instantaneous source term for the retained PCs also can be expressed in terms of a linear relation with the source terms for the thermo-chemical scalars:

$$\mathbf{s}_\phi = \mathbf{A}^\mathbf{T} \mathbf{s}_\theta \tag{11}$$

Therefore, knowledge of the chemical source terms of the thermo-chemical scalars can be used to reconstruct the PCs chemical source terms; and the same matrix relation applies to conditional means for these source terms (i.e. $\langle \mathbf{s}_\phi \rangle = \mathbf{A}^\mathbf{T} \langle \mathbf{s}_\theta \rangle$) given that the matrix $\mathbf{A}^\mathbf{T}$ is constant.

### 2.3. Joint PDF construction via KDE

As stated earlier, a second ingredient for the development of closure is associated with the joint scalar PDF. In contrast to presumed PDF shapes, the desired joint PDF model must have the following criteria: 1) it can be constructed based on data, which corresponds in the present framework to multi-scalar point measurements for thermo-chemical scalars, and 2) it must accommodate a wide range of shapes and dimensions. The dimension of the constructed PDF corresponds to the number of PCs retained to map the composition space.

In this work, the construction of joint PDFs is carried out using a kernel density estimation (KDE) method. KDE is a non-parametric way to estimate the PDF of a random variable. A $d$-dimensional kernel density function (KDF) is expressed as a sum of kernel functions centered on model determined points, which are learned from sample data [19]:

$$P(\boldsymbol{\phi}; \widetilde{\boldsymbol{\phi}}) = \frac{1}{nh}\sum_{i=1}^{n} K\left(\frac{\boldsymbol{\phi}-\widehat{\boldsymbol{\phi}}_i}{h}\right) \tag{12}$$

where, $K$ is the kernel function, $h$ is the bandwidth (smoothing) and $\widehat{\boldsymbol{\phi}}_i$ are the $n$ samples of the selected PCs for a given condition, $\widetilde{\boldsymbol{\phi}}$. $\widetilde{\boldsymbol{\phi}}$ is the Favre average of a PC vector $\boldsymbol{\phi}$ at a given position in space; although, if the same averages are available at different positions, a single joint PDF is constructed for them. A common kernel function, which is used in this study, is the Gaussian distribution.

It is noteworthy that in addition to the Favre mean of the PCs, we can, in principle, use higher moments (e.g. variances) to further parameterize the PDF. This invariably also requires the transport of these moments, if needed and additional closure for these moments. The chemical source terms for these moments can be derived from the instantaneous statistics. However, by comparisons of the reconstruction of thermo-chemical scalar moments in the flames studied indicate that the Favre means are sufficient. We attribute this to the uncorrelated nature of the PCs; a vector of averaged PCs indicates a relatively unique set of statistics in a given data set.

The bandwidth is an important parameter and determines the amount of smoothing of the kernel function. Larger bandwidths tend to smooth the data and are less likely to capture all the trends present in the distribution. The joint PDFs constructed in this case may resemble Gaussian or Beta distributions. On the other hand, very small bandwidths result in highly fluctuating functions. The optimal bandwidth can be determined through several estimation techniques [30]. For the present study, we are determining this bandwidth by trial-and-error 1) based on the number of single shot measurements at any given position and 2) by selecting the small bandwidth that maintains a smooth PDF in any given direction. We have observed that bandwidths ranging between 0.005 and 0.05 for a Gaussian kernel function are enough to smoothly capture the important characteristics present in the experimental data. The joint KDE is implemented in Python using the in-built functions available in their *scikit-learn* library [31]. These joint PDFs are constructed at each axial and radial location in the flame using the instantaneous pointwise measurements.

## 3. Results and discussion

In this section, we present results of an *a priori* analysis implemented on the Sandia flames D, E and F data. This analysis pertains to the determination of parameters, the PCs, which can construct the conditional statistics and the joint PDFs. The constructed statistics are compared with statistics evaluated on the raw data.

### *3.1. Run conditions*

The Sandia flames D, E and F are piloted non-premixed flames with similar inlet compositions and temperature and different flow conditions. The configuration consists of a round fuel jet (with an inner diameter of 7.2 mm) surrounded by a round pilot burner (of inner and outer diameters of 7.7 and 18.2 mm, respectively) which is also surrounded by co-flow air. The fuel is a 75% $CH_4$ and 25% air mixture. The pilot composition is based on the unstrained premixed flame solution of a methane-air flame at an equivalence ratio of 0.88 and at a temperature of 1880 K. The fuel jet Reynolds numbers based on the jet diameters are 22,400, 33,600 and 44,800 for flames D, E and F, respectively. The temperatures at the inlet of the fuel, pilot and co-flow streams are 294 K, 1880 K, 291 K, respectively. The ambient pressure is 0.993 atm.

In the experimental data of Sandia flames D, E and F, there are 300-1000 instantaneous measurements available at each measurement location. The size of the database, its sparsity in composition space, the magnitude of the experimental uncertainty and the dimension of the PDF and conditional means tables are all factors that can determine the adequacy of this database for a particular application. Since the proposed framework is a data-based framework, coordination with the experiment may be crucial towards the development of models that can adequately reconstruct the different model elements.

The Favre and conditional averages obtained using the PCA-KDE approach are compared with experimental Favre and conditional averages for Sandia flames D, E and F at several axial locations such as $x/d$ = 7.5, 15, 30 and 45. Most of the turbulent chemistry interaction phenomenon like local extinction, re-ignition etc. occurs at the axial distances mentioned above and, hence, are chosen for comparison. Here, comparisons are shown for temperature and mass fractions of OH, $H_2$ and CO. The other major variables like $O_2$, $H_2O$, $CO_2$ and $CH_4$ are highly correlated with temperature and have similar accuracies as temperature. In this work, PCA is carried out using custom subroutines implemented in Python.

*3.2. Conditioning parameters*

As stated earlier, in this work a PCA is carried out on all the instantaneous measurements available from Sandia flames D, E and F [21] at different axial and radial locations. The thermo-chemical scalars used in the PCA include $T$, $H_2$, $O_2$, OH, $H_2O$, $CH_4$, CO and $CO_2$. Figure 1 shows a scree plot of the percentage variance captured by each principal component. The PCs are ordered based on their contributions to the total variance in the source data.

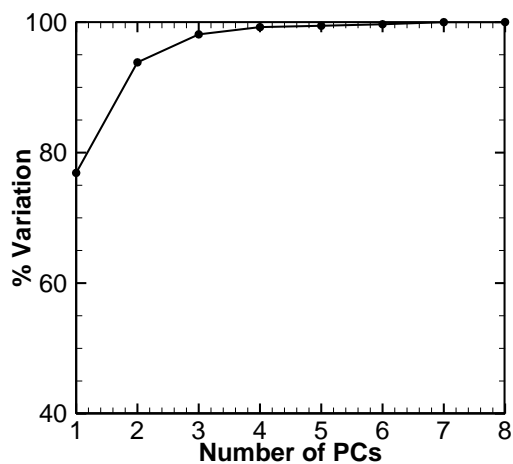

**Figure 1.** Scree plot: % Variation vs. Number of PCs.

It may be observed from Fig. 1 that 2 PCs capture about 95% of the total variation in the data and, hence, can be used to represent the set of thermo-chemical variables. Therefore, a dimensional reduction from 8 measured thermo-chemical scalars to 2 PCs is established. One extra PC may be retained to implement the model within the context of RANS or LES (i.e. in a *a posteriori* simulation) to establish solution accuracy as the model is integrated in space and time.

An important question arises regarding the physical meaning of the PCs. Based on Eq. (6), individual PCs can be expressed as a linear combination of the normalized thermo-chemical scalars. Much of the common variables used to parameterize the composition space are often expressed as a linear combination of such scalars, including the mixture fraction or the reaction progress variable. Since the Sandia flames are already well studied and well understood flames, it would be interesting to attempt to relate the established PCs to the common parameters that are used to parameterize their composition space.

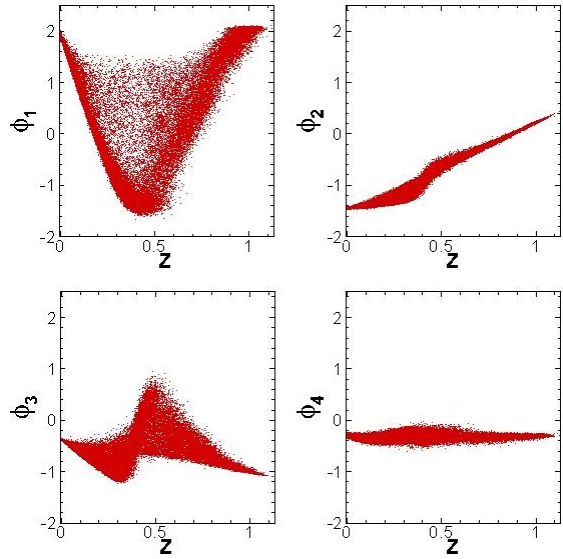

**Figure 2.** Correlations of the leading 4 PCs with mixture fraction for flame F.

Figure 2 shows scatter plots of the leading 4 PCs vs. the mixture fraction in flame F. The first PC exhibits trends in mixture fraction space that are complementary to those of the temperature or a progress variable with peaks at pure fuel or pure oxidizer conditions are minima around the stoichiometric conditions (mixture fraction of approximately 0.35). It is important to note that if a parameter is a PC, then, any variable that can be expressed in terms of a linear relation not that PC can also serve as a PC. The second PC is linear with the mixture fraction at fuel-rich and fuel-lean conditions away from the reaction zone. Therefore, the second PC serves essentially as a substitute for a mixture fraction. The third and fourth PCs may be best interpreted from the values of the coefficients of the eigenvector matrix $Q$. We re-write Eq. (9) explicitly in terms of the contributing thermo-chemical scalars in the Sandia flames as follows:

$$\begin{pmatrix} \phi_1 \\ \phi_2 \\ \phi_3 \\ \phi_4 \\ \phi_5 \\ \phi_6 \\ \phi_7 \\ \phi_8 \end{pmatrix} = \begin{pmatrix} \mathbf{-0.514} & -0.098 & \mathbf{0.508} & -0.237 & \mathbf{-0.428} & 0.127 & -0.172 & \mathbf{-0.429} \\ -0.042 & 0.226 & -0.219 & -0.077 & 0.026 & \mathbf{0.899} & 0.277 & -0.088 \\ 0.074 & \mathbf{-0.464} & 0.019 & \mathbf{-0.473} & -0.100 & 0.374 & \mathbf{-0.595} & 0.226 \\ 0.259 & -0.139 & -0.062 & \mathbf{-0.834} & 0.021 & 0.098 & -0.366 & 0.264 \\ 0.159 & \mathbf{-0.781} & 0.034 & -0.089 & -0.120 & 0.025 & \mathbf{0.584} & -0.028 \\ -0.325 & 0.135 & -0.080 & 0.006 & \mathbf{-0.555} & -0.043 & 0.197 & \mathbf{0.722} \\ \mathbf{0.665} & 0.220 & -0.016 & 0.082 & \mathbf{-0.663} & -0.020 & -0.021 & -0.249 \\ 0.297 & 0.220 & \mathbf{0.826} & 0.058 & 0.212 & 0.153 & 0.153 & 0.322 \end{pmatrix} \begin{pmatrix} T^* \\ Y^*_{H_2} \\ Y^*_{O_2} \\ Y^*_{OH} \\ Y^*_{H_2O} \\ Y^*_{CH_4} \\ Y^*_{CO} \\ Y^*_{CO_2} \end{pmatrix} \quad (13)$$

Note that the superscript "*" indicates a normalization of the thermo-chemical scalars using minimum and maximum values in the data such that their values range from -1 to 1. The normalization is designed to allow for equal contributions from all quantities measured. Also, we are highlighting values of the coefficients in the $\mathbf{A^T}$ matrix with magnitudes greater than 0.4. The choice of the highlighted number is arbitrary; but it helps in identifying the larger contributors to the PCs from the normalized thermo-chemical scalars. The various contributions of the thermo-chemical scalars to the PCs can be identified by considering each row of $\mathbf{A^T}$ aligned with those PCs. The first row indicates that the primary contributions to the $PC_1$ may be attributed to either temperature, reactants or products. Therefore, it represents a reaction progress variable as suggested by Fig. 2. The dominant contribution to $PC_2$ is associated with the fuel, which is highly correlated with the mixture fraction for mixture fractions above the stoichiometric value. Parente et al. [28] also identified a strong correlation of one of the PCs with the mixture fraction in $CH_4/H_2$ flames propagating into a hot and diluted co-flow at different levels of oxygen dilution. A second contribution to that PC comes from the intermediates $H_2$ and CO and the oxidizer, with the later also correlated with the mixture fraction for values of the mixture fraction below the stoichiometric value. For the third PC, the contribution is mainly dominated by intermediates $H_2$, CO and OH. Therefore, it represents a measure of the degree of chemical reactivity. The fourth PC is primarily correlated with OH. As a minor species, compared to the other intermediates on the list, $H_2$ and CO, it represents finer structures in the reaction zone; and, therefore, $PC_4$ appears to be the specialized PC to resolve it. Although, there are other quantities that also have non-negligible contributions to its value.

Based on the above observations, it is clear that the third and the fourth PCs can play an important role in predicting OH in addition to the first two PCs.

In the following discussion, we carry out *a priori* analysis of the reconstruction of thermo-chemical scalars statistics (conditional means and PC joint PDFs) by relying only on the first 2 PCs. However, in Sec. 3.5, we also include results based on the retention of 4 PCs to compare measured scalars' radial statistics at different downstream distances from the jet inlet.

### *3.3. Construction of the conditional means*

Having identified the leading PCs, means of thermo-chemical scalars can be constructed conditioned upon the selected PCs. Figure 3 shows the conditional means of temperature, OH, $H_2$, CO, $CO_2$, $H_2O$, $CH_4$ and $O_2$ mass fractions conditioned on the first two PCs, $\phi_1$ and $\phi_2$. The statistics are constructed using a binning procedure from all instantaneous measurements from the Sandia flame D, E and F [21] based the corresponding values of $\phi_1$ and $\phi_2$. One hundred bins are considered across each PC's range of values. Empty bins resulting from the sparsity of data in higher dimensions are approximated by linear interpolation involving neighboring populated bins. We have found that empty bins in PC space can account for less than 3% of all bins within the accessed composition space if 2 PCs are used. Increasing the number of retained PCs requires further coarse-graining of the binning procedure (i.e. using a lower number of bins for each PC's range).

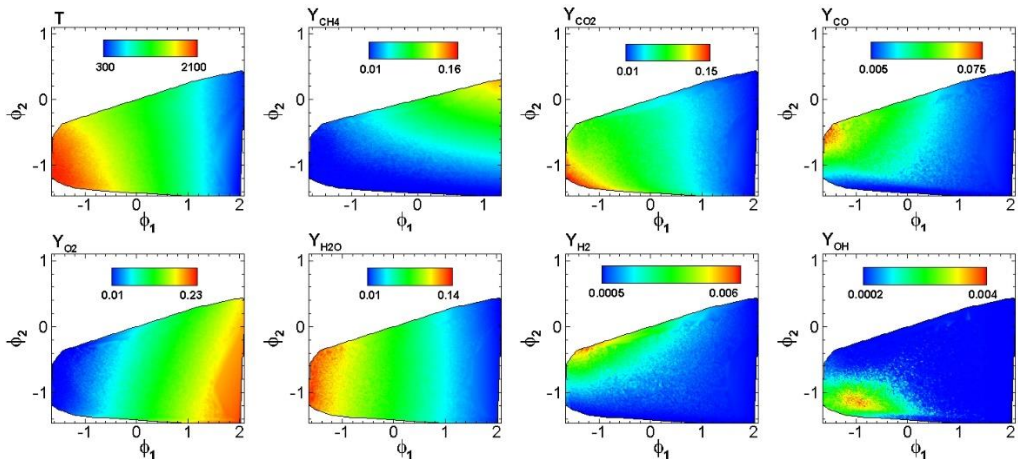

**Figure 3.** Conditional statistics of measured species and temperature.

### *3.4. Construction of the joint PDFs*

Now, having identified the PCs, we investigate the construction of the joint PDFs and compare their predictions based on the raw data or by use of the KDE construction. With dimensionality reduction, the joint PDF based on several thermo-chemical scalars can be replaced with a joint PDF based on just the two PCs:

$$P(T, Y_{H_2}, Y_{O_2}, Y_{OH}, Y_{H_2O}, Y_{CH_4}, Y_{CO}, Y_{CO_2}) \rightarrow P(\phi_1, \phi_2; \tilde{\phi}_1, \tilde{\phi}_2) \qquad (14)$$

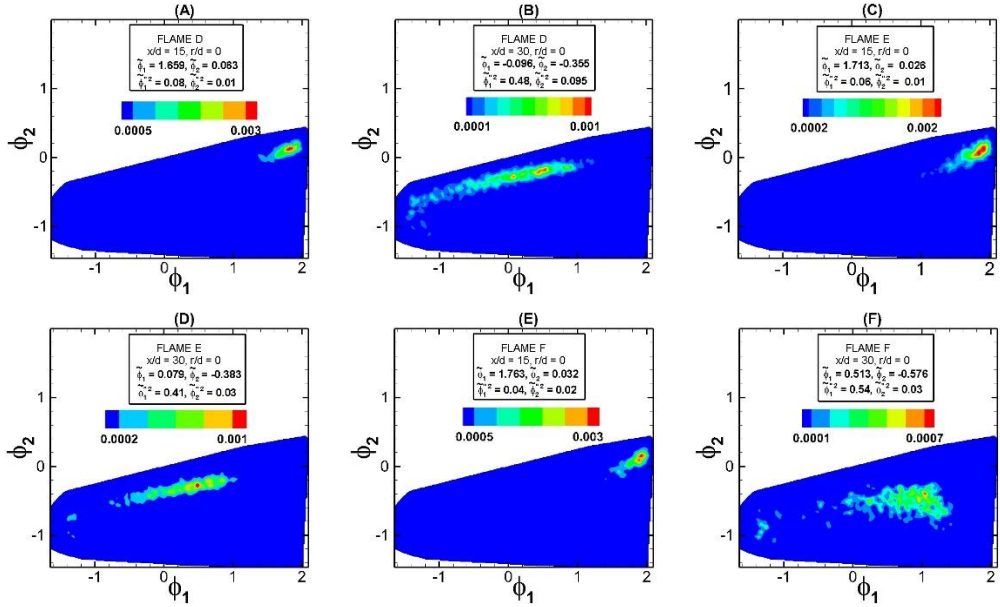

**Figure 4.** Joint PDFs for flame D (A, B), flame E (C, D) and flame F (E, F) at $r/d = 0$ and $x/d = 15, 30$.

In this work, PDFs conditioned on PCs, $\phi_1$ and $\phi_2$ are constructed using the instantaneous point measurements for flames D, E and F at all radial and axial locations where measurements were carried out. One key advantage of the KDE approach relative to a histogram-based construction of the joint PDF is the ability of KDE to establish smoother distributions even for a limited set of data. As stated earlier, joint PDFs are constructed at other radial and axial location of the three flames. Then, a tabulation can be constructed to establish the PDFs for a wide range of Favre-averaged values for the pairs of the PCs. The effectiveness of these PDFs depends on their ability to recover smooth mean and RMS statistics that match well with experimental data. Although not shown here, characterizing the PDFs with variances of the PCs in addition to their Favre-means does not improve the comparisons of the radial profiles of the mean and RMS values of the thermo-chemical scalars discussed below.

Figure 4 shows the KDE-based joint PDFs with respect to PCs, $\phi_1$ and $\phi_2$ constructed for flames D, E and F at the centerline ($r/d = 0$) and two different downstream distances, $x/d = 15$ and 30. The downstream distance of $x/d = 15$ corresponds approximately to the onset of extinction for flames D, E and F; while, by $x/d = 30$, the reignition process is being completed. The figure shows that while significant similarities in the PDF are exhibited at $x/d = 15$, there distributions tend to be broader for the higher Reynolds number flames at $x/d = 30$, indicating different rates of transition to the reignition process.

### *3.5. Favre averages and RMS statistics*

In this section, results from the construction of conditional means and joint PDFs based on the 2 retained leading PCs are combined to construct spatial statistics in terms of radial profiles at different downstream distances for flames D, E and F. The Favre-averaged mean and RMS statistics are computed by integrating the joint PDF with conditional statistics of the thermo-chemical variables as follows:

$$\widetilde{Y}_i = \frac{\int \langle \rho | \phi_1, \phi_2 \rangle \langle Y_i | \phi_1, \phi_2 \rangle\, P(\phi_1, \phi_2; \widetilde{\phi}_1, \widetilde{\phi}_2)}{\int \rho \langle \rho | \phi_1, \phi_2 \rangle\, P(\phi_1, \phi_2; \widetilde{\phi}_1, \widetilde{\phi}_2)} \tag{15}$$

$$\widetilde{Y_i''^2} = \frac{\int \langle \rho | \phi_1, \phi_2 \rangle \langle Y_i^2 | \phi_1, \phi_2 \rangle\, P(\phi_1, \phi_2; \widetilde{\phi}_1, \widetilde{\phi}_2)}{\int \langle \rho | \phi_1, \phi_2 \rangle\, P(\phi_1, \phi_2; \widetilde{\phi}_1, \widetilde{\phi}_2)} - \widetilde{Y_i}^2 \tag{16}$$

where, $P$ is the joint PDF, $\rho$ is conditional density and $Y_i$ are conditional statistics of $i^{th}$ thermo-chemical scalars. The conditional statistics of each thermo-chemical variables are generated on a 2-D, $\phi_1$-$\phi_2$, space by linearly interpolating from an ensemble of instantaneous measurements of flames D, E and F.

Figure 5 shows comparisons between the PCA-KDE approach and the direct evaluation from experimental data of the Favre averages and RMS statistics of temperature and CO, $H_2$ and OH mass fractions for flame D at axial locations $x/d = 7.5, 15, 30$ and $45$. The direct evaluation from experimental data is based on carrying out Favre averages of the thermo-chemical scalars at the positions of the experimental measurements at a given $x/d$. The figure shows that the PCA-KDE approach matches well with experimental statistics for flame D. The joint PDFs constructed in reduced dimensions can recover all the trends and peaks of the experimental data at different axial locations.

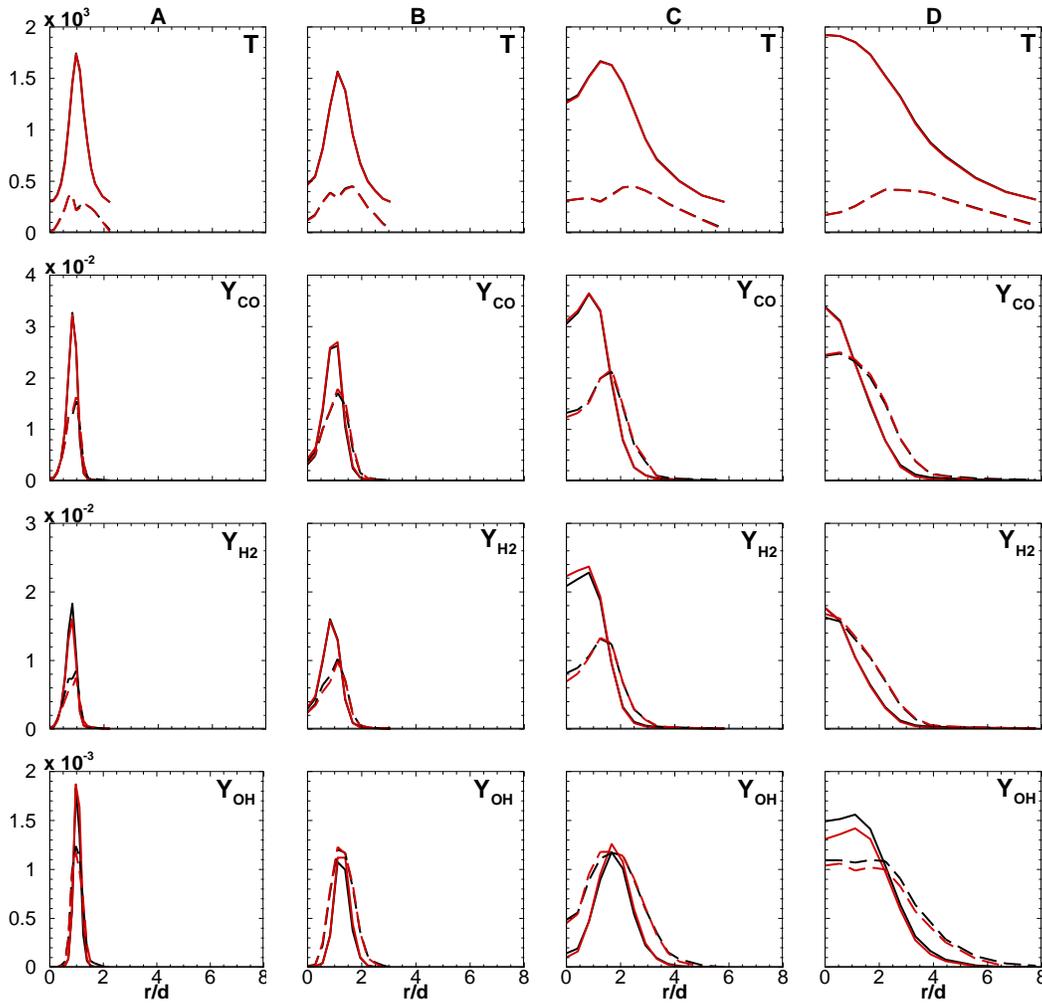

**Figure 5.** Flame D Favre averages (solid) and RMS (dashed) comparisons of experimental data (red) and PCA-KDE with 2 PCs (black) at A) $x/d = 7.5$, B) $x/d = 15$, C) $x/d = 30$ and D) $x/d = 45$.

Next, we present similar plots as in Fig. 5 for flames E (Fig. 6) and flame F (Fig. 7) at the same axial locations, $x/d = 7.5, 15, 30$ and $45$. Starting from flame D local extinction becomes increasingly visible as the Reynolds number is increased. The PCA-KDE is able to track effectively these trends. It is important to note that the conditional means used for the evaluation of the PCA-KDE-based statistics are the same for all flames. Therefore, the difference in the trends are exhibited by the differences in the joint PC PDFs, which reflect the increasing rate of non-equilibrium effects in flames E and F. Local extinction reaches a peak at an axial distance of about 15-30 jet radii. These trends are clearly exhibited by the lower peaks of the intermediate's CO and $H_2$.

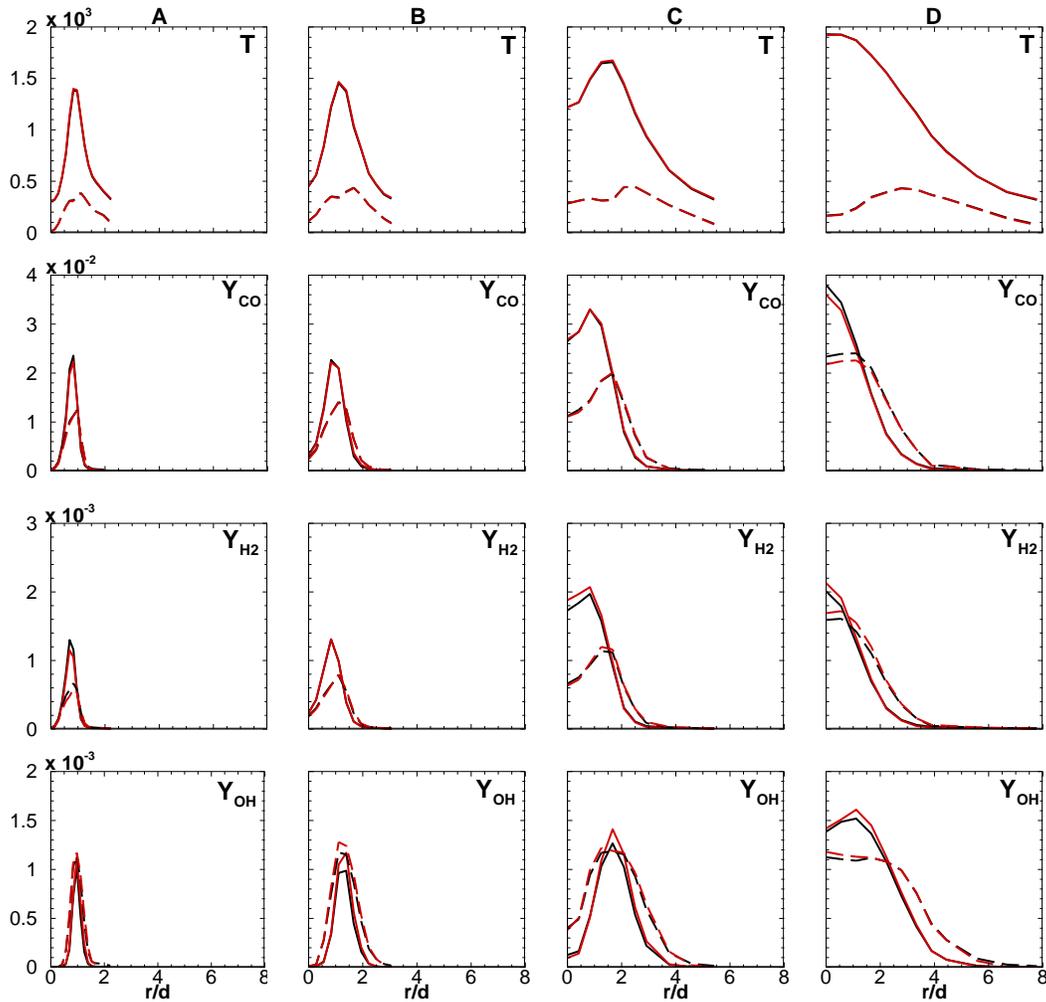

**Figure 6.** Flame E Favre averages (solid) and RMS (dashed) comparisons of experimental data (red) and PCA-KDE with 2 PCs (black) at A) $x/d = 7.5$, B) $x/d = 15$, C) $x/d = 30$ and D) $x/d = 45$.

The results from Figs. 5-7 clearly indicate the validity of the proposed PCA-KDE framework based on t parameterization of the composition space in terms of a limited number of PCs. Therefore, if PCs are transported, then, an adequate reconstruction of the thermo-chemical scalars' statistics can be carried ou solution in a post-processing step.

At this stage it is important to reiterate that the joint PDF formulation and the mean and RMS statistics' reconstruction have been accomplished using 2 PCs. As discussed in previous sections, additional PCs may be required to fully represent the composition space and capture all the fine-grained structures in the flame. The challenge of using additional PCs lies in the construction of multi-dimensional joint PDF as well as the resulting sparsity in the conditional statistics. Although, not shown here, the mean and RMS averages exhibit slight improvements with 3 PCs. However, as we go to 4 or more PCs, we are limited by the amount of data that can span the inherent multidimensionality of the statistics. With 4 PCs and despite the value-added contributions of the 4$^{th}$ PC, we have to use a coarse-grained strategy to represent our statistics (e.g. using 20 bins for each PC range instead of 100 bins to determine conditional means) with no discernable improvements in the radial statistics.

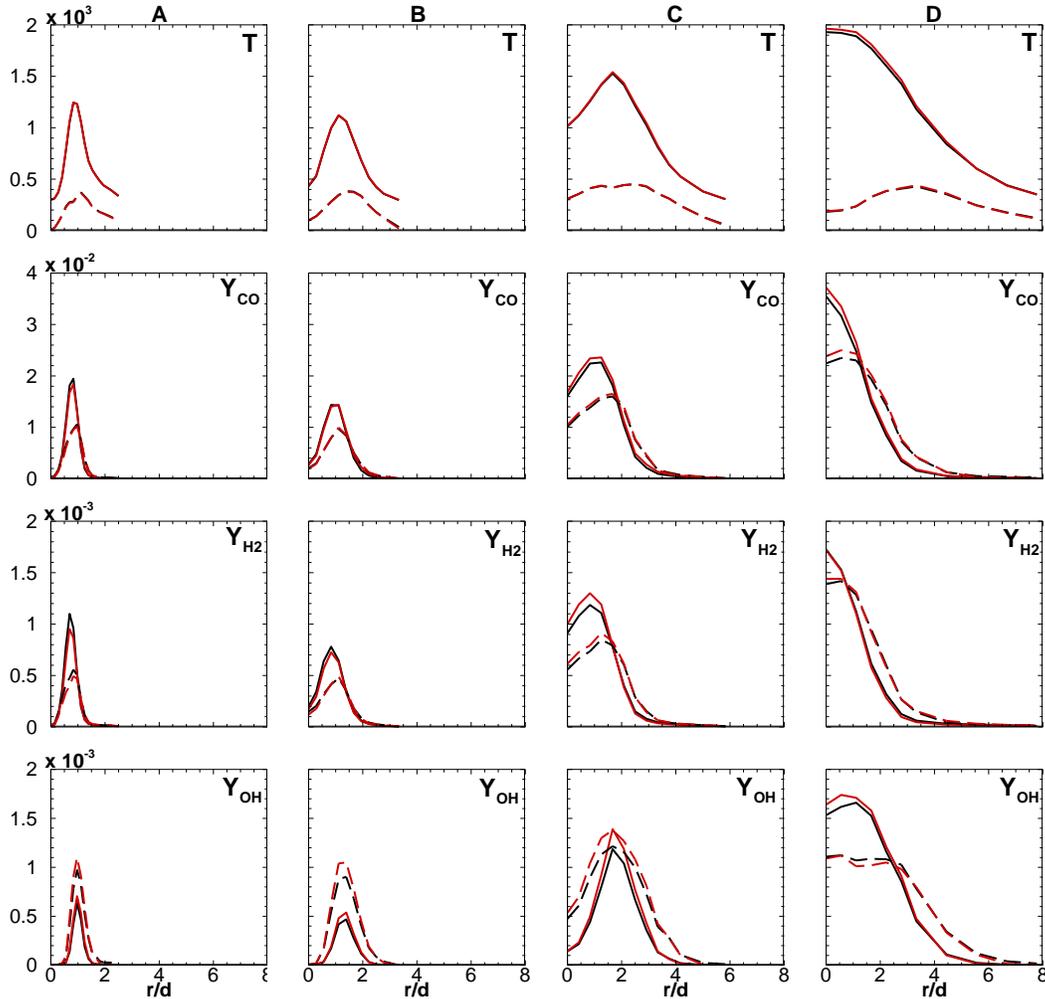

**Figure 7.** Flame F Favre averages (solid) and RMS (dashed) comparisons of experimental data (red) and PCA-KDE with 2 PCs (black) at A) $x/d = 7.5$, B) $x/d = 15$, C) $x/d = 30$ and D) $x/d = 45$.

Overcoming the limitations of data sparsity requires a close coordination with the experimentalists who generate the data. Nonetheless, we also are currently investigating different strategies to improve the prediction of conditional means and joint PDFs with higher dimensional parameterizations. For example, we are currently looking at implementing independent component analysis (ICA) [32,33], a similar method to PCA designed to project parameterized data in terms of statistically independent components (ICs). ICA can be implemented on the retained PC by carrying out a rotation of these PCs (i.e. a multiplication by a constant matrix) to establish ICs. Similar to PCA, source terms for ICs can be determined from the measured thermo-chemical scalars' source terms through a linear relation. With statistically independent ICs to parameterize the composition space, the multi-dimensional joint PDF can be expressed as the product of the individual ICs marginal PDFs, thus, reducing the requirements on the measurement database size. Similarly, for the conditional means, the PMSR sets can be expanded beyond their "local mixing" region to include data from a larger region in the composition space to spawn new data. This new data serves as a substitute for the "filling" the composition space with interpolation.

The framework described here requires 1) a model for the PCs chemical source terms, and 2) an inversion of the PC solution to the thermo-chemical scalars' solution. Such a model would be constructed after carrying out the *a priori* steps that are described here, which involve 1) an identification of the PCs that parameterize the composition space using PCA, 2) a tabulation of the conditional means and joint PDFs based on these PCs, 3) the development of a model for the means of the PCs chemical source terms, and 4) an evaluation of an inversion relation between the PC and thermo-chemical scalars.

At this stage, it is interesting to address an important question related to whether the presence of mapping from PCs to thermo-chemical scalars and *vice versa* based on the instantaneous variables can be used to approximate the same mapping between Favre-averaged means of the PCs and the thermo-chemical scalars. The question is related to the inherent similarities between the governing equations for PCs and thermo-chemical scalars in their instantaneous and averaged forms when differential diffusion effects between effects are not significant compared to the role of turbulent transport (i.e. the turbulent diffusivity).

If a similarity between the mapping relations exists, then, the mapping can provide an alternative to the PCA-KDE reconstruction process such that the Favre-averaged statistics of the thermo-chemical scalars can be determined directly from the transported PCs using the same mapping relations between the instantaneous PCs and the instantaneous thermo-chemical scalars. In our present study, we use artificial neural networks (ANN) to reconstruct instantaneous thermo-chemical scalars from a smaller number of PCs. We have found that ANN's application as a regression tool can be effective in recovering thermo-chemical scalars from PCs [24]. To investigate the PC-to-thermo-chemical scalars' mapping, we consider two different network architectures based on 2 and 4 PCs for each measured thermo-chemical scalar. Both ANN architectures are based on a single hidden layer with 40 neurons but differ in the number of inputs based on the number of PCs used. The goal is to examine the required number of PCs to accurately estimate statistics of all measured scalars.

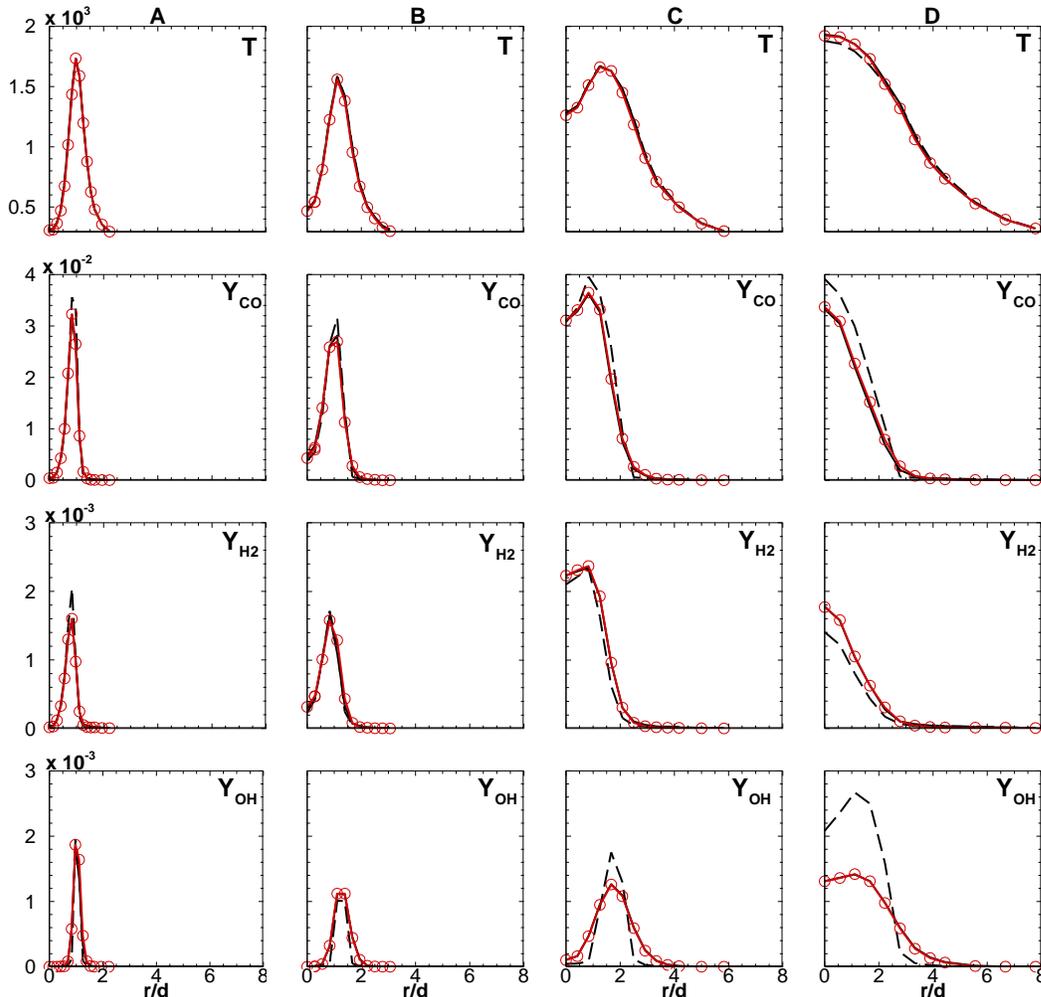

**Figure 8.** Comparison of Flame D Favre averages with 2 PCs (dashed), 4 PCs (solid) using ANN-inversion and experimental data (red) at A) $x/d = 7.5$, B) $x/d = 15$, C) $x/d = 30$ and D) $x/d = 45$.

Figures 8-10 compares the Favre-averaged scalar profiles for temperature and CO, $H_2$ and OH mass fractions at various axial locations obtained from the ANN-inversion method with 2 and 4 PCs and experimental data (shown in red and symbols to distinguish the profiles from the mapping relations). The inversion is carried out by 1) evaluating the Favre-averaged PCs at each measurement point, then, 2) using ANN to look up the thermo-chemical scalars at the same positions. The comparisons are made for all flames D, E and F. From the figures, it is clear that for the case of 4 PCs, the comparison with experiment shows an excellent agreement for the temperature and the selected species, including OH. Although not shown here, other major species like $H_2O$, $CO_2$, $CH_4$ and $O_2$ are also captured well by the ANN inversion procedure using both 2 PCs and 4 PCs. With 2 PCs, significant differences can be observed in the radial profiles of OH at axial distances of $x/d = 30$ and 45. Regardless, the ANN inversion relations can provide an alternative to the PCA-KDE tabulation to estimate thermo-chemical scalars mean profiles from mean PC profiles.

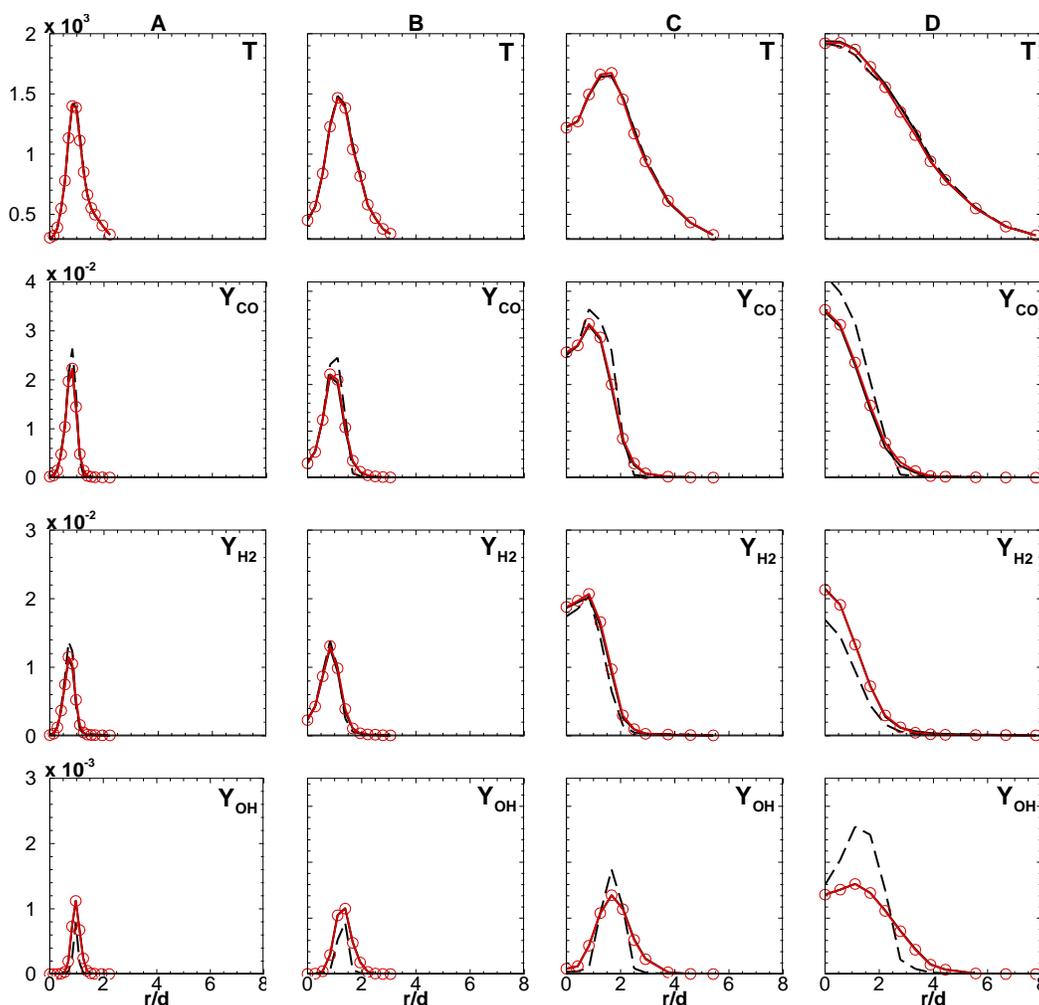

**Figure 9.** Comparison of Flame E Favre averages with 2 PCs (dashed), 4 PCs (solid) using ANN-inversion and experimental data (red and symbols) at A) $x/d = 7.5$, B) $x/d = 15$, C) $x/d = 30$ and D) $x/d = 45$.

Based on the above observations, the inversion from PCs to thermo-chemical scalars, which can be developed from instantaneous data, can be used to estimate the Favre-averaged thermo-chemical scalars from Favre-averaged PCs. Such an inversion can provide an independent assessment of the evaluation of the Favre-averaged thermo-chemical scalars from its tabulation based on Eq. (9).

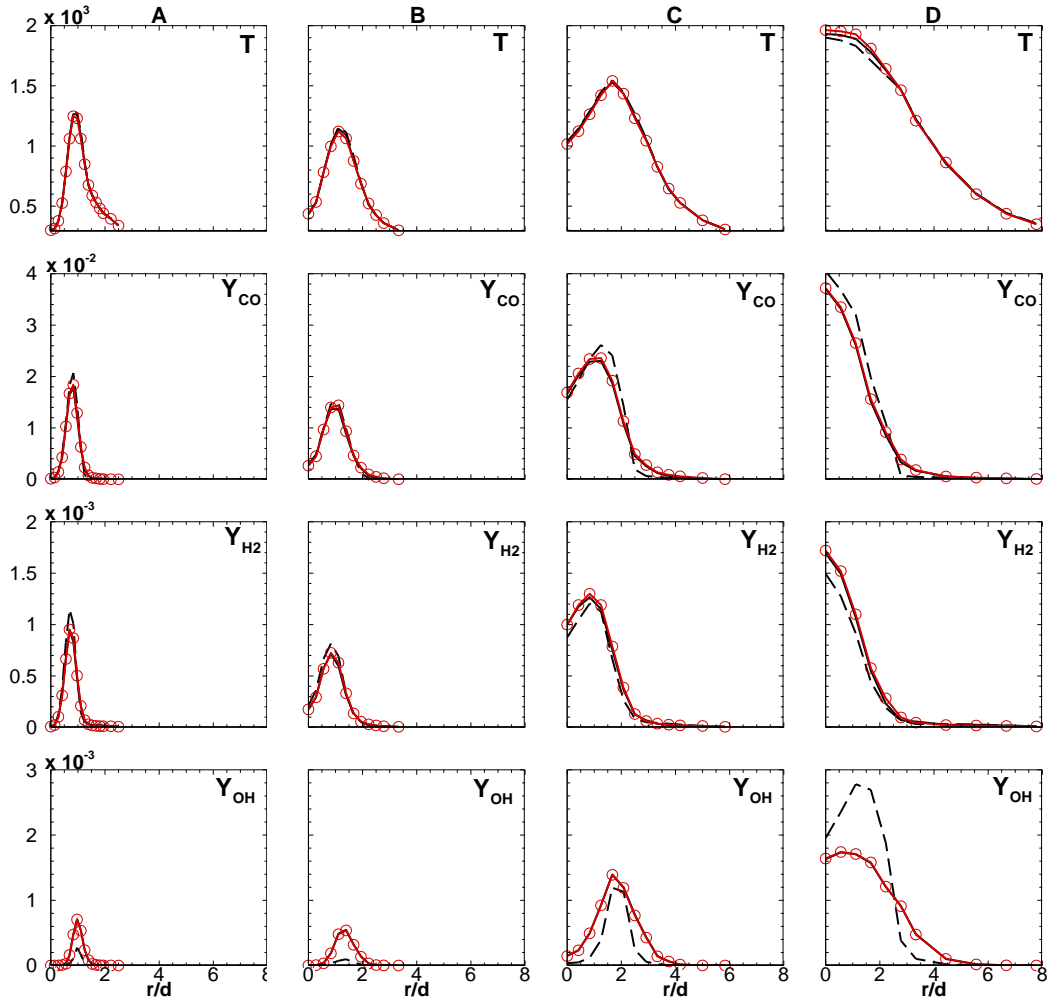

**Figure 10.** Comparison of Flame F Favre averages with 2 PCs (dashed), 4 PCs (solid) using ANN-inversion and experimental data (red and symbols) at A) $x/d = 7.5$, B) $x/d = 15$, C) $x/d = 30$ and D) $x/d = 45$.

### *3.6. Validation of chemical source term reconstruction methodology*

Since experimental data does not provide any information on the species source terms, we carry out an *a priori* analysis of the approach presented in Sec. 2.2 for recovering species chemical source terms using numerical simulation data. The data is based on one dimensional turbulence (ODT) simulations of the Sandia flames. ODT is a one-dimensional model that treats the coupling between reaction and molecular transport deterministically while capturing the 3D effects of stirring stochastically [34]. In the past, the second author has implemented ODT for the study of co-flowing jet flames [35] and more specifically for the study of the Sandia flames in [36-37]. The model formulation for a temporal jet flame can be found in [33]. The model is based on the 1D unsteady solutions for the streamwise momentum, energy and species equations. The temporal evolution can be converted to a downstream evolution of the 1D profiles using a parabolic approximation [35]. The sequence of random number selections for implementing the stirring events can generate different realizations of the same ODT solutions to obtain statistics, which can be constructed at each 1D spatial location and temporal evolution. To validate the model, we also introduce "experimental noise" to the instantaneous ODT data to emulate experimental measurements. The experimental error model is prescribed as follows:

$$\phi_{\text{perturbed}} = \phi_{\text{simulated}} + 0.02 \times r \times \phi_{max}, \qquad (16)$$

where $r$ is a random function ranging from -1 to 1 and $\phi$ corresponds to the thermo-chemical scalars that are chosen to be the measured scalars. These scalars are identical to the ones measured in the experimental data. A 2% error is chosen to remain consistent with the experimental uncertainty.

For testing purposes, ODT simulations are carried out for Sandia flame D to generate instantaneous data at different axial (interpreted from the temporal evolution of 1D profiles) and radial locations (along the 1D direction). This numerical data contains source term information of temperature and measured species and is a good source for validation of the source term reconstruction. As a starting point to the PMSR-based reconstruction algorithm, the temperature and measured species are initialized with the ODT perturbed values while the rest of the unmeasured species are set to zero. The modified PMSR run is carried out with the procedure described earlier in Sec. 2.2 to recover these unmeasured species and eventually construct source terms at every measurement point. Then, the unperturbed ODT results and the PMSR reconstructed results are compared in PC space. The same binning procedure as above is adopted for constructing the conditional means based on 2 PCs. Figure 11 shows the comparison between conditional source means from PMSR and the source terms of the original unperturbed ODT data for *T*, $O_2$, $H_2O$, $CH_4$ and $CO_2$. Figure 12 shows similar contours for $H_2$, CO and OH. Also shown are contours of the error for some of the quantities drawn with different scales from conditional means. The source terms for the species are in kg/m$^3$-sec and for the temperature, they are in kJ/m$^3$-sec.

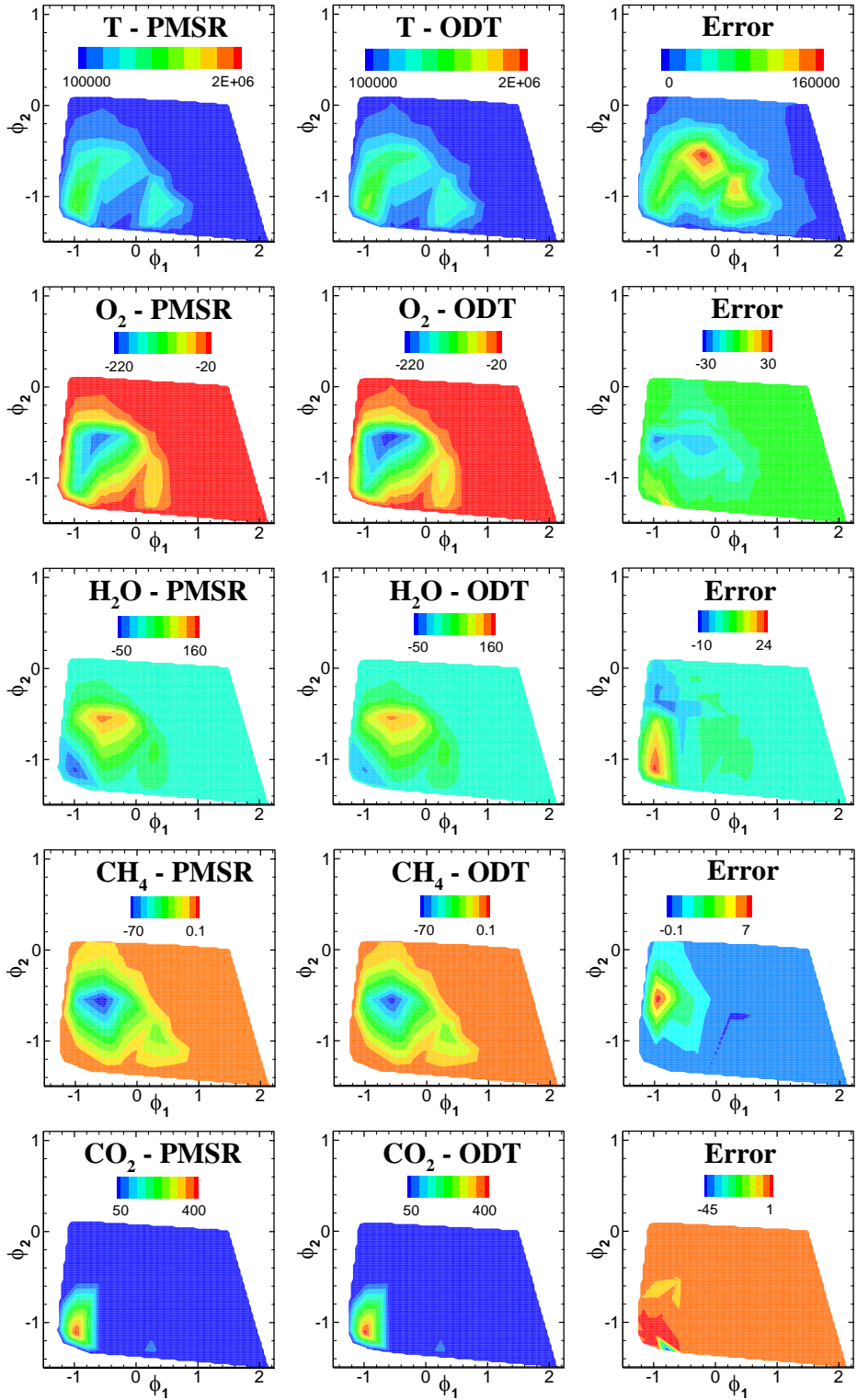

**Figure 11.** Conditional source term comparison of PMSR reconstruction (left) and ODT (middle) for $T$, O2, H$_2$O, CH$_4$ and CO$_2$.

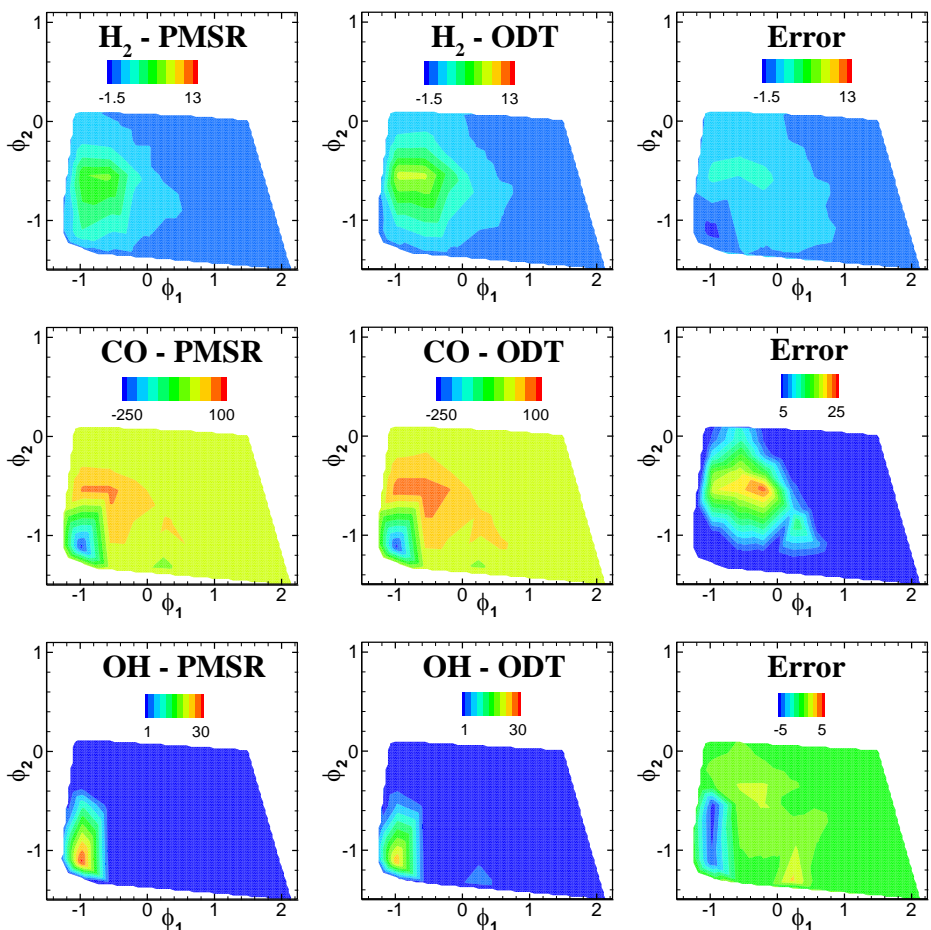

**Figure 12.** Conditional source term comparison of PMSR reconstruction (left) and ODT (middle) for $H_2$, CO and OH.

Figures 11 and 12 show that the conditional source terms obtained from reconstruction using PMSR match reasonably well with those of ODT. The broader trends including the magnitude and position of peaks are captured reasonably well. Although not shown here, the chemical source terms for these thermo-chemical scalars based on the initial conditions of the PMSR (perturbed measured quantities and zero values for unmeasured quantities) are not even comparable to the statistics based on the perturbed ODT data. This may be attributed to the nonlinear form of the reaction rates, which can amplify the measurement error and the lack of key radicals and intermediates, which play an important role in the fuel oxidation stages. Regardless, further work is needed to construct accurate conditional means for the chemical source terms of the thermo-chemical scalars and the PCs. We are presently investigating strategies to "de-noise" the experimental data using deep machine learning techniques, which have been successful in other applications with noisy observations [38]. The implementation of such strategies will be critical for accurately implementing the framework and the solution of the governing equations (1)-(3).

## 4. Conclusion

In this paper, we present a novel framework for developing closure models in turbulent combustion based on multi-scalar measurements. The closure relies on the construction of parameters that describe the composition space and its statistics based on PCs. KDE is used to construct multi-variate statistics conditioned on the first moments of the PCs. These PCs have a set of important properties, which make them particularly useful in combustion problems. First, they provide a generic set of variables that can be used to parameterize the composition space, especially when such parameters cannot be prescribed in advance. Also, because such parameters represent an uncorrelated set, further reduction in the description of their distributions can be carried out, potentially resulting in a lower number of equations needed to advance the solutions for the PC-based governing equations in RANS or LES.

The PCA-KDE approach is validated on experimental data for the Sandia flames D, E and F. The approach presented here can construct smooth joint PDFs that capture the key characteristics of the three flames amidst noisy and limited data sets. The mean and RMS statistics match reasonably well with the experimental data indicating that the constructed joint PDF can predict effects of local extinction and re-ignition. The conditional averages based on all three flames match well with experimental data.

Although, data from point measurements can be readily available to extract PDFs, such data can potentially be interpreted to estimate filtered density functions (FDFs), which are useful within the context of LES. Moreover, the strategy adopted for PDF (or potentially FDF) construction can be incorporated within the context of PDF transport methods by serving as an alternative strategy for constructing the PDFs based on the transport of notional particles.

Finally, this work is only the first step towards implementing a data-based closure model for turbulent combustion. One of the key implementations for these models is the development of closure for the PCs chemical source terms. We have presented and demonstrated one strategy of extracting these chemical source terms from experimental measurements using a modified PMSR approach. We are investigating other strategies to de-noise the experimental data using machine learning techniques to improve these predictions.